\documentclass[aps,twocolumn,nofootinbib,superscriptaddress,PRD]{revtex4}
\usepackage{amsmath,bm}
\usepackage{graphicx}
\usepackage{epsfig}	
\usepackage{multirow}
\usepackage{color}
\usepackage{xcolor}
\begin{document}

\newcommand{\vk}{{\vec k}}
\newcommand{\vK}{{\vec K}}
\newcommand{\vb}{{\vec b}}
\newcommand{{\vp}}{{\vec p}}
\newcommand{{\vq}}{{\vec q}}
\newcommand{\vQ}{{\vec Q}}
\newcommand{\vx}{{\vec x}}
\newcommand{\beq}{\begin{equation}}
\newcommand{\eeq}{\end{equation}}
\newcommand{\half}{{\textstyle \frac{1}{2}}}
\newcommand{\gton}{\stackrel{>}{\sim}}
\newcommand{\lton}{\mathrel{\lower.9ex \hbox{$\stackrel{\displaystyle<}{\sim}$}}}
\newcommand{\ee}{\end{equation}}
\newcommand{\ben}{\begin{enumerate}}
\newcommand{\een}{\end{enumerate}}
\newcommand{\bit}{\begin{itemize}}
\newcommand{\eit}{\end{itemize}}
\newcommand{\bc}{\begin{center}}
\newcommand{\ec}{\end{center}}
\newcommand{\bea}{\begin{eqnarray}}
\newcommand{\eea}{\end{eqnarray}}

\newcommand{\beqar}{\begin{eqnarray}}
\newcommand{\eeqar}[1]{\label{#1} \end{eqnarray}}
\newcommand{\pleft}{\stackrel{\leftarrow}{\partial}}
\newcommand{\pright}{\stackrel{\rightarrow}{\partial}}

\newcommand{\eq}[1]{Eq.~(\ref{#1})}
\newcommand{\fig}[1]{Fig.~\ref{#1}}
\newcommand{\eff}{ef\!f}
\newcommand{\alphas}{\alpha_s}

\renewcommand{\topfraction}{0.85}
\renewcommand{\textfraction}{0.1}
\renewcommand{\floatpagefraction}{0.75}

\title{Exposing the dead-cone effect of jet quenching in QCD medium}

\date{\today  \hspace{1ex}}

\author{Yun-Fan Liu}
\affiliation{School of Mathematics and Physics, China University
of Geosciences, Wuhan 430074, China}

\author{Wei Dai}
\email{weidai@cug.edu.cn}
\affiliation{School of Mathematics and Physics, China University
of Geosciences, Wuhan 430074, China}

\author{Ben-Wei Zhang}
\email{bwzhang@mail.ccnu.edu.cn}
\affiliation{Key Laboratory of Quark \& Lepton Physics (MOE) and Institute of Particle Physics, Central China Normal University, Wuhan 430079, China}
\affiliation{Guangdong Provincial Key Laboratory of Nuclear Science, Institute of Quantum Matter, South China Normal University, Guangzhou 510006, China}

\author{Enke Wang}
\affiliation{Guangdong Provincial Key Laboratory of Nuclear Science, Institute of Quantum Matter, South China Normal University, Guangzhou 510006, China}
\affiliation{Guangdong-Hong Kong Joint Laboratory of Quantum Matter, Southern Nuclear Science Computing Center, South China Normal University, Guangzhou 510006, China}
\affiliation{Key Laboratory of Quark \& Lepton Physics (MOE) and Institute of Particle Physics, Central China Normal University, Wuhan 430079, China}

\begin{abstract} 

When an energetic parton traverses the hot QCD medium it may suffer multiple scattering and lose its energy. The medium-induced gluon radiation for a massive quark will be suppressed relative to that of a light quark due to the dead-cone effect. The development of new declustering techniques of jet evolution makes a direct study of the dead-cone effect in the QCD medium possible for the first time. In this work, we compute the emission angle distribution of the charm-quark initiated splittings in $\rm D^0$ meson tagged jet and that of the light parton initiated splittings in an inclusive jet in p+p and Pb+Pb at $\rm 5.02$~TeV by utilizing the declustering techniques of jet evolution. The heavy quark propagation and indued energy loss in the QCD medium are simulated with the SHELL model based on the Langevin equation. By directly comparing the emission angle distributions of charm-quark-initiated splittings with those of light parton-initiated splittings at the same energy intervals of the initial parton, we provide insights into the fundamental splitting structure in A+A collisions, thereby exploring the possible observation of the dead-cone effect in medium-induced radiation. We further investigate the case of the emission angle distributions normalized to the number of splittings and find the dead-cone effect will broaden the emission angle of the splitting and reduce the possibility to occur such splitting, therefore leading the massive parton to lose less energy. We also find the collisional energy loss mechanism has a negligible impact on the medium modification to the emission angle distribution of the charm-quark initiated splittings for $\rm D^0$-meson tagged jets.



\end{abstract}


\maketitle

\section{Introduction}

In Quantum Chromodynamics (QCD), the vacuum induces a fast parton emitting gluon in a process that can be described as a parton shower. The parton shower evolves into a multi-parton final state, then harmonizes into a cluster of final state hadrons in a similar direction, which can be detected and recognized as jets~\cite{2019Looking}. It has been observed in particle colliders that the fast parton can be produced in the initial hard scatterings with large momentum transfer, and then make subsequent emissions resulting in additional productions of quarks and gluons, and then be reconstructed as jets in the final state.

One can expect the radiation from a parton of Mass ($\rm M$) and energy ($\rm E$) will be suppressed within an angular size of $\rm M/E$. Such a phenomenon was named as the dead-cone effect~\cite{Dokshitzer:1991fd}, which manifested itself indirectly in various heavy flavor-related observables in particle collider experiments~\cite{CMS:2019jis,DELPHI:2000edu,SLD:1999cuj,Llorente:2014bha}. An iterative jet declustering technique~\cite{Frye:2017yrw,Dreyer:2018nbf,Cunqueiro:2018jbh} has emerged to help expose the jet substructure to the most basic splitting structure experimentally.  ALICE report~\cite{ALICE:2021aqk} suggested a comparison between the emission angle distribution of charm-quark initiated splittings in $\rm D^0$-meson tagged jets and that of light parton initiated splittings in inclusive jets produced in p+p collision at $13$~TeV at proper radiator's energy intervals. One can directly observe the dead-cone effect of the charm quark for the first time.


In high-energy nuclear-nuclear collisions, the fast parton produced in the initial hard scattering may pass through the de-confined state of quark-gluon plasma (QGP). The medium modification of the fast parton is referred to as the jet quenching phenomenon~\cite{Wang:1992qdg,Gyulassy:2003mc,Qin:2015srf,Viinikainen:2023tji,Grosse-Oetringhaus:2011hoh} has been intensively studied theoretically and experimentally. One of the key mechanisms of jet quenching is expected to be the radiation of soft gluons induced by the scattering of the fast parton with the medium constituents~\cite{Guo:2000nz,Zhang:2003yn,Zhang:2003wk,Majumder:2009ge,Chen:2010te,Chen:2011vt,Vitev:2002vr,Arnold:2000dr,Arnold:2003zc,Baier:2000mf}. Such in-medium emission is also predicted to suffer the dead-cone effect which means the medium-induced radiation probability of the fast parton traversing the hot medium is also suppressed within an angular size of $\rm M/E$~\cite{Dokshitzer:2001zm,Zhang:2003wk,Armesto:2003jh,Zhang:2004qm}. It may result in several observations which are explained by the scenario of the heavy flavor quarks losing less energy than light flavor quarks in experimental measurements~\cite{PHENIX:2004ggw,STAR:2006btx,ALICE:2012ab,CMS:2016mah,ATLAS:2022fgb}. However, the dead-cone effect itself can not be isolated and observed through these measurements. In p+p collisions, iterative declustering techniques have allowed exposure of the dead-cone effect. The application of such techniques to jets in heavy ion collisions might allow us to expose the dead-cone effect in medium-induced radiation and to understand the mass dependency of jet quenching. In this paper, we calculate the emission angle distribution of the charm-quark initiated in-medium splittings in $\rm D^0$-meson tagged jets and that of the light parton initiated in-medium splittings in inclusive jets in Pb+Pb collisions at $\rm \sqrt{s}=5.02$~TeV, and search for expose the dead-cone effect of medium-induced radiation in jet quenching.

\section{Splitting-angle distributions in $\rm D^0$-meson tagged jets in p+p}

\begin{figure*}[htbp]

	\begin{center}
        \includegraphics[scale=0.8]{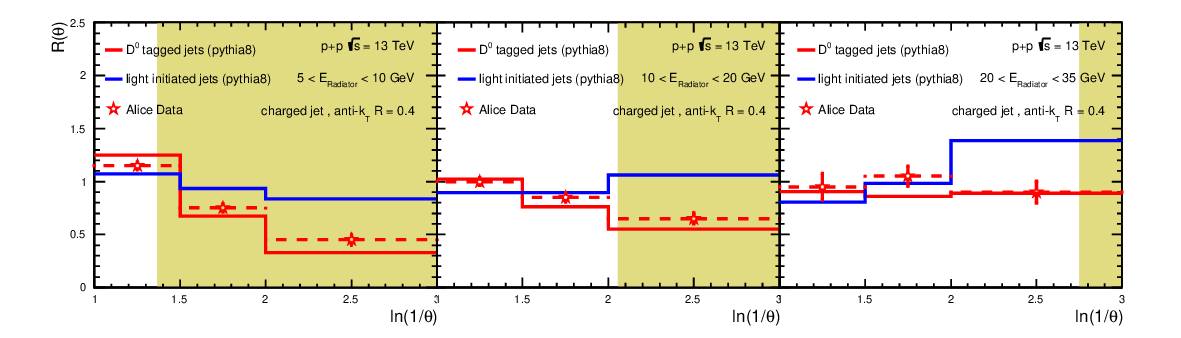}
	\end{center}
	\vspace{-0.03\textwidth}	
	\caption{The ratios of the splitting-angle distributions for $\rm D^0$-meson tagged jets (light-quark jets) to inclusive jets, $\rm R(\theta)$ in p+p collisions at $\rm \sqrt{s}=13$~TeV using PYTHIA $\rm 8$. The results are demonstrated for three energy intervals of the radiators: $\rm 5<E_\text{Radiator}<10$~GeV (left panel),  $\rm 10<E_\text{Radiator}<20$~GeV (middle panel) and  $\rm 20<E_\text{Radiator}<35$~GeV (right panel) and they are compared with ALICE experimental data. The shaded areas correspond to the angles in which the radiation is suppressed due to the dead-cone effect, the mass of the charm quark is assumed to be $1.275$~GeV/$\rm c^2$.}
	\label{fig:13tevbaseline}
\end{figure*}

In this section, we reproduce the exact ALICE’s setup and observable in pp collisions to validate our vacuum reference. The observable is defined as:
\begin{eqnarray}
\rm R(\theta)=\frac{1}{N^{D^0 jets}} \frac{dn^{D^0jets}}{dln(1/\theta)} / \frac{1}{N^{inclusive\ jet}} \frac{dn^{inclusive\ jet}}{dln(1/ \theta)},
\label{observable}
\end{eqnarray}
where the numerator is the splitting-angle distributions of charm-quark initiated splittings for $\rm D^0$-meson tagged jets normalized to the number of jets, and the denominator represents the splitting-angle distributions for inclusive jets also normalized to the number of jets. The ratio is taken for the same initial energy $\rm E_{Radiator}$ intervals. The no dead-cone limit of such observable can be derived from calculating the splitting-angle distribution in pure light-initiated jets from event generators and then replacing the numerator in Eq.~\ref{observable} to be $\rm (1/N^{lightjets})(dn^{lightjets}/dln(1/\theta))$.

PYTHIA 8~\cite{Sjostrand:2007gs} is used as the event generator, the anti-$k_T$ algorithm~\cite{Cacciari:2008gp} from the Fastjet package~\cite{Cacciari:2011ma} is used to simulate the production of jets with a transverse momentum in the interval of $\rm 5< p^{ch}_{T, jet}<50$~GeV. The $\rm D^0$-meson selected in the transverse-momentum interval $\rm 2< p^{D^0}_T<36$~GeV/c. Once the required jets are selected, the internal splitting tree is reconstructed. After the preparation of the jets, in the iterative de-clustering processes, the Cambridge-Aachen (C/A) algorithm~\cite{Dokshitzer:1997in} is implemented to recluster the constituents in jets using the angular-ordered nature of the splittings, the same as the QCD emissions.  By iterative declustering the splitting tree, the reclustering history is exposed. The measurements of the primary two prongs structures are recorded at each declustering step: the angle between the daughter prongs in splittings, $\theta$, the relative momentum transfer of the splitting, $\rm k_T$, and the energy of the parton initiating the splitting (the radiator), $\rm E_{Radiator}$.  $\rm k_T > \Lambda_{QCD}=200 $~MeV$/c$ is implemented to suppress the hadronisation effects~\cite{Lifson:2020gua}. The prongs containing $\rm D^0$-mesons are always being followed to reconstruct the gluon emission history off a charm quark in a vacuum with an emission angle at each splitting process. The contaminations of gluon splitting contribution of the heavy flavor production were studied with MC simulations and found to be negligible. A minimum cut is imposed to the leading charged hadron in the leading prong of the recorded splitting in inclusive jets, $\rm p^{leadinghadron}_{T} \geq 2.8$~GeV$/s$ which corresponds to the lower transverse momentum cut $2$~GeV of the $\rm D^0$-meson in $\rm D^0$-meson tagged jets.

\begin{figure*}[htbp]
	\begin{center}
        \includegraphics[scale=0.8]{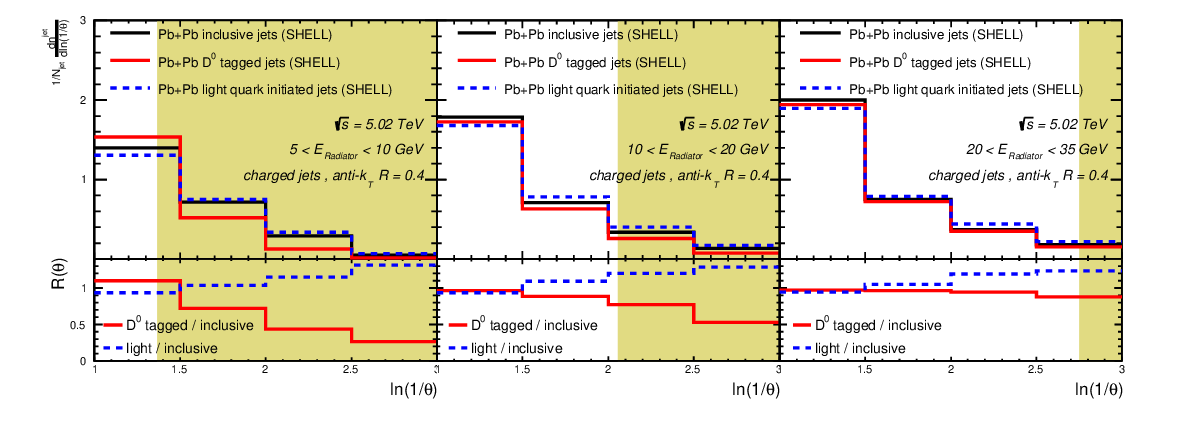}
	\end{center}
	\vspace{-0.03\textwidth}	
	\caption{The splitting-angle distributions for $\rm D^0$-meson tagged jets, inclusive jets and also light-quark jets normalized to the number of jets in Pb+Pb collisions at $\rm \sqrt{s}=5.02$~TeV (upper plots) and also the $D^0$-meson tagged jets/inclusive jets (light-quark jets/inclusive jets) ratios (bottom plots) calculated for three energy intervals of the radiators: $\rm 5<E_\text{Radiator}<10$~GeV (left panel),  $\rm 10<E_\text{Radiator}<20$~GeV (middle panel) and  $\rm 20<E_\text{Radiator}<35$~GeV (right panel). The shaded areas correspond to the angles at which the radiation is suppressed due to the dead-cone effect.}
	\label{fig:dcinAA}
\end{figure*}

In Fig.~\ref{fig:13tevbaseline}, we reproduce the ratios of splitting-angle distributions $\rm R(\theta)$ for $\rm D^0$-meson tagged jets (light-quark jets) to inclusive jets in p+p collisions at $\rm \sqrt{s}=13$~TeV using PYTHIA $\rm 8$. The results are demonstrated for three energy intervals of the radiators: $\rm 5<E_\text{Radiator}<10$~GeV (left panel),  $\rm 10<E_\text{Radiator}<20$~GeV (middle panel) and  $\rm 20<E_\text{Radiator}<35$~GeV (right panel) which is as same as Ref.~\cite{ALICE:2021aqk} in order to compare with ALICE data respectively. In these comparisons, PYTHIA $\rm 8$ is proven to be sufficient to fairly describe the differences between the heavy and light quark-initiated splitting angular distributions within a jet along with the analysis procedures mentioned above. The observable $\rm R(\theta)$ do can help demonstrate the differences of the radiation angular distributions of a charm quark and light quarks, then further expose the dead-cone effect of charm quark by illustrating suppression in the $\rm \ln(1/\theta)$ regions which are colored in each $\rm E_\text{Radiator}$ interval. These colored areas are corresponding to the dead-cone angles in each interval, $\rm \theta_{dc} < m_c/E_\text{Radiator}$. With the increase of $\rm E_\text{Radiator}$, we can observe the suppression in the lower $\theta$ region is weaker because the dead-cone region of $\theta$ is smaller.

When understanding the `suppression,' theoretically, $\rm R(\theta)$ can not compare with unity since $\rm N^{D^0 jets}$ and $\rm N^{inclusive\ jet}$ will not be the same, the denominator of such ratio that is used to compare with the $\theta$ distributions in  $D^0$-meson tagged jets are reconstructed from inclusive jets in which there are light-quark, and gluon initiated splittings involved.  Still, the advantage of such an observable is that it is experimentally measurable and still decisively manifests the suppression in the dead-cone region of $\theta$. Moreover, the plots with no dead-cone limits are also plotted accordingly.
The deviations between the $\rm R(\theta)$ curve of $\rm D^0$-meson tagged jets and the no dead-cone limit are much more pronounced.

\section{Theoretical Framework of the in-medium Evolution of $D^0$-meson tagged jets in A+A}

When a fast parton traverses the hot and dense medium, it will lose its energy by radiating gluons due to multiple scatterings in such medium. The spectrum of the radiative gluon is calculated from multiple gluon emission theories~\cite{Guo:2000nz,Zhang:2003yn,Zhang:2003wk,Majumder:2009ge,Chen:2010te,Chen:2011vt,Vitev:2002vr,Arnold:2000dr,Arnold:2003zc,Baier:2000mf}. In this letter, we implement the formalism of the Higher-Twist approach ~\cite{Guo:2000nz,Zhang:2003yn,Zhang:2003wk}:

\begin{eqnarray}
\rm \frac{dN}{dxdk^{2}_{\perp}dt}=\frac{2\alpha_{s}C_sP(x)\hat{q}}{\pi k^{4}_{\perp}}\sin^2(\frac{t-t_i}{2\tau_f})(\frac{k^2_{\perp}}{k^2_{\perp}+x^2M^2})^4\ ,
\label{eq:dNdxdk2}
\end{eqnarray}
where $\rm x$ and $\rm k_\perp$ are the energy fraction and transverse momentum of a radiated gluon, $\rm M$ is the mass of the parent parton, $\rm \alphas$ is the strong coupling constant, $\rm C_s$ is the quadratic Casimir in color representation, $\rm P(x)$ is the vacuum splitting function~\cite{Wang:2009qb}, the jet transport coefficient $\rm \hat{q} \propto \hat{q}_0(T/T_0)^3$~\cite{Chen:2010te}, $\rm \tau_f = 2Ex(1-x)/(k_{\perp}^2 + x^2M^2)$ is the gluon formation time. 

The dead-cone effect manifests itself in the last quartic term of Eq.~(\ref{eq:dNdxdk2}).  One can easily rewrite such term to be: $(1+\theta_0^2/\theta^2)^{-4}$ with the relation $\theta_0 = M/E$ and $k_\perp = x E\theta$. The radiative gluon spectrum for massive quarks will be largely suppressed when the radiation angle $\theta < \theta_0$.  It is very interesting to investigate how the detailed effect manifests itself in the QCD medium-modified emission angle distributions of heavy quark-initiated splittings.

\begin{figure*}[htbp]
    \begin{center}
        \includegraphics[scale=0.8]{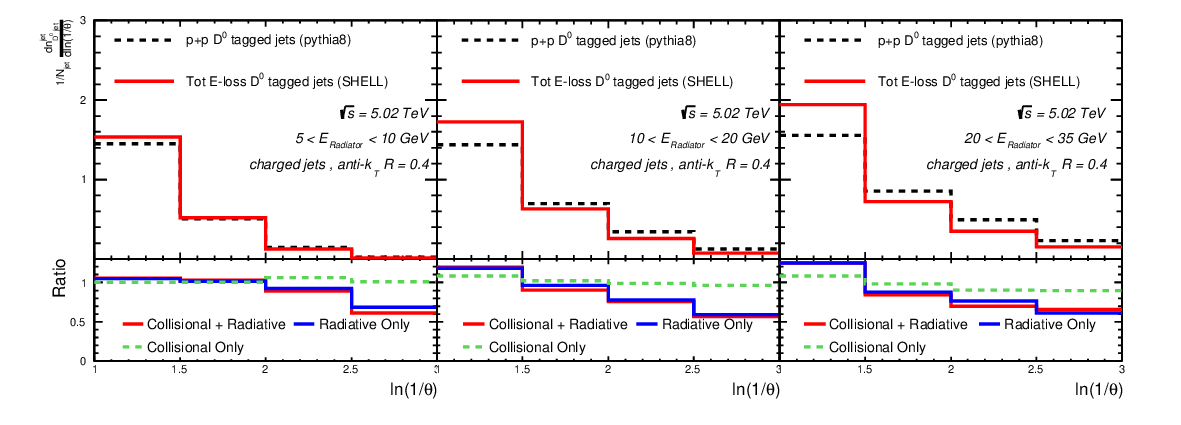}
	\end{center}
	\vspace{-0.05\textwidth}	
	\caption{The splitting-angle distributions of the $\rm D^0$ meson tagged jets normalized to the number of jets in p+p and A+A collisions at $\rm \sqrt{s}=5.02$~TeV (upper plots)  and also their A+A/p+p ratios (bottom plots) calculated for three energy intervals of the radiators: $\rm 5<E_\text{Radiator}<10$~GeV (left panel),  $\rm 10<E_\text{Radiator}<20$~GeV (middle panel) and  $\rm 20<E_\text{Radiator}<35$~GeV (right panel). Also, the contributions from the radiative and collisional energy losses in A+A collisions are presented accordingly. }
	\label{fig:raaheavy}
\end{figure*}

\begin{figure*}[htbp]
\begin{center}
        \includegraphics[scale=0.8]{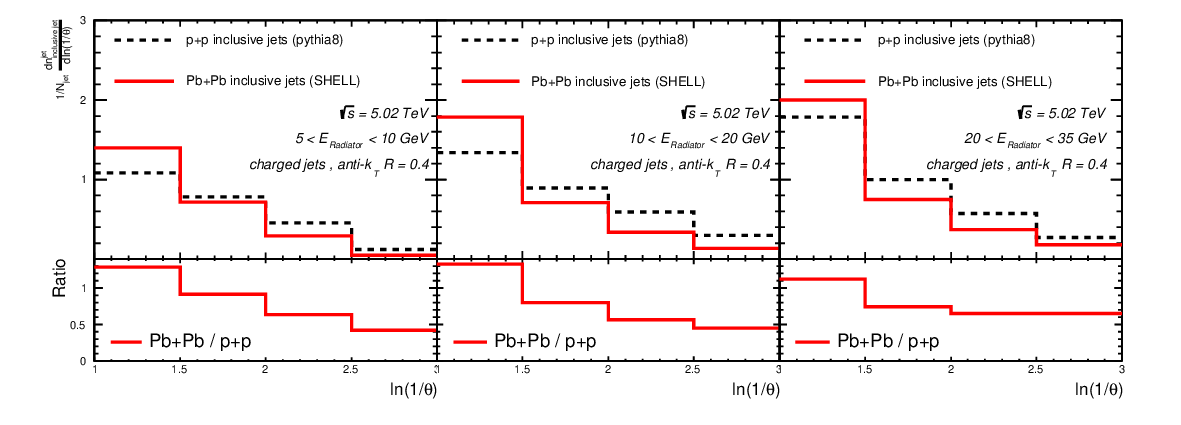}
	\end{center}
	\vspace{-0.05\textwidth}	
	\caption{The splitting-angle distributions of the inclusive jets normalized to the number of jets in p+p (dash lines) and A+A (solid lines) collisions at $\rm \sqrt{s}=5.02$~TeV (upper plots) and also their A+A/p+p ratios (bottom plots) calculated for three energy intervals of the radiators: $\rm 5<E_\text{Radiator}<10$~GeV (middle panel),  $\rm 10<E_\text{Radiator}<20$~GeV (right panel) and  $\rm 20<E_\text{Radiator}<35$~GeV (left panel)}
	\label{fig:raainclusive}
\end{figure*}

Taking advantage of the previous study in p+p collisions, we compute the same charm quark-initiated splittings and light parton-initiated splittings to demonstrate the possible dead-cone effect in A+A collisions. The Simulating Heavy quark Energy Loss with Langevin equations (SHELL) model~\cite{Dai:2018mhw,Wang:2020ukj,Wang:2020bqz,Wang:2020qwe} is employed to describe the in-medium evolution of the heavy parton considering both collisional and radiative energy loss mechanism. The SHELL model has already been utilized to predict the number of heavy flavor jet-related observable~\cite{ATLAS:2022fgb,CMS:2019jis,ALICE:2021njq}. It is built on the framework of the modified Langevin equations describing the propagation of heavy quarks~\cite{Cao:2013ita,Dai:2018mhw,Wang:2019xey,Wang:2020bqz,Wang:2020qwe},
\begin{eqnarray}
\label{eq:lang1}
\rm 
&&\Delta\vec{x}(t) = \frac{\vec{p}(t)}{E}\Delta t \ ,\\
&&\Delta\vec{p}(t) = -\Gamma(p,T)\vec{p}\Delta t+\vec{\xi}(t)\sqrt{\Delta t}-\vec{p}_g(t) \, ,
\label{eq:lang2}
\end{eqnarray}
where $\Delta t$ is the time interval between each Monte Carlo simulation step, drag coefficient $\rm \Gamma$ is constrained by the fluctuation-dissipation relation~\cite{He:2013zua} with momentum diffusion coefficient $\rm \kappa$, $\rm  \Gamma =\kappa/2ET=T/D_s E$ and $\rm D_s =4/2\pi T$. $\rm D_s$ is the spacial diffusion coefficient. The thermal stochastic term $\vec{\xi}(t)$, which is random kicks on the heavy quarks from thermal quasi-particles in QGP, obeys the Gaussian distribution $\rm \left \langle \xi^i(t)\xi^j(t') \right \rangle =\kappa \delta^{ij}\delta(t-t')$. The last term in Eq.~(\ref{eq:lang2}) is the momentum recoil due to the medium-induced gluon radiation, which is implemented with the higher-twist approach mentioned in Eq.~\ref{eq:dNdxdk2}.
During each time interval, the in-medium gluon radiation probability, which is also calculated from Eq.~\ref{eq:dNdxdk2} and assumed to obey the Poisson distribution, is implemented to decide whether radiation happens during a Langevin evolution time interval, $\rm P(n)=\lambda^n e^{-\lambda}/n!$ means the probability $\rm P(n)$ of radiating $n$ gluons during such time interval $\rm \Delta t$. $\rm \lambda$ is the mean value of $n$ and can be calculated by integrating Eq.~(\ref{eq:dNdxdk2}),
\begin{eqnarray}
\rm \lambda(t,\Delta t)=\Delta t\int dxdk^2_{\perp} \frac{dN}{dxdk^2_{\perp} dt}
\label{eq:lambda}
\end{eqnarray} 
If radiation occurs, the number of radiated gluons is determined by $P(n)$ and the four-momentum of each radiated gluon can be updated in Eq.~(\ref{eq:dNdxdk2}), \textit{i}.\textit{e}., the last term $\rm \vec{p}_g$ of Eq.~(\ref{eq:lang2}) in each time interval $\rm \Delta t$. The simulation of a parton propagating in the hot and dense medium will keep evolving as described above until the temperature of the medium decreases to $T_c = 165$~MeV. The space-time evolution of the QCD medium is provided by a (3+1)D viscous hydrodynamic model CLVisc~\cite{Pang:2012he}. An energy cutoff $\rm \omega_0 \ge \mu_D = \sqrt{4\pi\alphas}T$ is set to maintain the fluctuation-dissipation relation for heavy quarks, and the initial parton positions are provided by Glauber Monte Carlo~\cite{Miller:2007ri}. The value of $\rm \hat{q}_0 = 1.5\ \rm{GeV^2/fm}$ is extracted from $\chi^2$ calculations in final-state hadron productions in Pb + Pb collisions at $\rm \sqrt{s} $ = 5.02~TeV~\cite{Ma:2018swx,Zhang:2022fau}. The value $\rm D_s(2\pi T) = 4$ extracted from $\chi^2$ calculations in $D$ meson $\rm R_{AA}$ between theoretical calculations and experimental data~\cite{STAR:2014wif,Xie:2016iwq,CMS:2017qjw,ALICE:2018lyv} is consistent with $\rm D_s(2\pi T) = 3.7 \sim 7$ obtained from Lattice QCD calculations~\cite{Francis:2015daa,Brambilla:2020siz}.

The radiative energy loss of light partons is considered the same as that of heavy quarks in the same Langevin time step. To simulate the collisional energy loss of the light partons, a Hard Thermal Loop (HTL) formula\cite{Neufeld:2010xi} has been adopted in this work:
$\rm \frac{dE^{coll}}{dt}=\frac{\alpha_{s}C_{s}\mu^{2}_{D}}{2}ln\frac{\sqrt{ET}}{\mu_{D}}$.
When all the partons stop their propagation in QGP medium and fragment into hadron, we first construct strings using the colorless method developed by the JETSCAPE collaboration\cite{Putschke:2019yrg}, then perform hadronization and hadron decays using the PYTHIA Lund string method.

\section{Observation of the dead-cone in A+A collisions}

In Fig.~\ref{fig:dcinAA}, we calculate the splitting-angle distributions normalized to the number of jets for both $\rm D^0$-meson tagged jets and inclusive jets respectively in Pb+Pb collisions at $\rm \sqrt{s}=5.02$~TeV displayed in the upper panels, denoted as solid lines in different colors. They are presented in the same three energy intervals of the radiators as in p+p, furthermore, the results for light flavor initiated jets case are also plotted as no-quark mass (dead-cone) reference. Then the ratio of these two distributions is expected to help expose the possible suppression in the dead-cone region of splitting-angle $\rm \theta$.  By comparing the $\rm R^{AA}(\theta)$ distributions as a function of $\rm \ln{(1/\theta)}$ with the no quark mass (dead-cone) limit at the different radiators' energy intervals, we find such suppression begins to vanish when the energy of the radiator increases.  However, there will also be a mass effect in the collisional energy loss mechanism which is not expected to be affected by such suppression in the dead-cone region of the splitting angle. It is important to investigate such pollution to isolate the observation of the dead-cone effect implemented in the radiative energy loss mechanism.

To verify such suppression is truly due to the dead-cone effect embedded in the formalism of the higher-twist approach that describes the radiative energy loss due to multiple scattering, we compute in the upper panels of Fig.~\ref{fig:raaheavy} the splitting-angle distributions of the $\rm D^0$ meson tagged jets normalized to the number of jets in Pb+Pb collisions at $\rm \sqrt{s}=5.02$~TeV. Also, the A+A/p+p ratios from separated contributions for collisional energy loss and radiative energy loss in the bottom panel are denoted as the dashed line and solid lines respectively.  We find that the radiative energy loss contribution would lead the distribution in Pb+Pb to shift to a larger $\rm \theta$ region than p+p for all three energy intervals of radiators. However, the collisional energy loss contribution would barely affect the $\rm \ln{(1/\theta)}$ distributions for the $\rm D^0$ meson-tagged jets in p+p. We can conclude that the collisional energy loss mechanism has a negligible impact on the medium modification to the emission angle distribution of the charm-quark initiated splittings for $\rm D^0$ meson-tagged jets. Only the radiative energy loss contribution will be responsible for the medium modification to the $\rm \ln{(1/\theta)}$ distributions for the $\rm D^0$ meson-tagged jets. Therefore, $\rm R_{AA}(\theta)$  can be computed using the same iterative jet declustering techniques applied in p+p and then measured experimentally. It also can be proposed to disclose the measurement of the dead-cone effect of the medium-induced radiation in jet quenching. It will further help constrain the radiative energy loss description of jet quenching models and help gain more understanding of the in-medium evolution of the jet shower.

\begin{table}
    \centering
    \begin{tabular}{|c|c|c|c|}
        \hline
        \multirow{2}{*}{ $E_\text{Radiator}$} & Inclusive jets  & $\rm D^0$ jets  &  \multirow{2}{*}{} \\
        \cline{2-3}
         &   $\rm \langle \theta \rangle_{jets}$  &   $\rm \langle \theta \rangle_{jets}$    & \\
         \hline
        \multirow{2}{*}{$\rm 5-10$~GeV} & 0.31 & 0.34 & pp \\
        \cline{2-4}
         & 0.36 & 0.36 &AA\\
         \hline
           \multirow{2}{*}{$\rm 10-20$~GeV} & 0.40 & 0.37 & pp \\
        \cline{2-4}
         & 0.45 & 0.42 &AA\\
         \hline
          \multirow{2}{*}{ $\rm 20-35$~GeV} & 0.47 & 0.42 & pp \\
        \cline{2-4}
         & 0.49 & 0.47 &AA\\
         \hline
    \end{tabular}
    \caption{The averaged splitting-angles in jets for $\rm D^0$-meson tagged and inclusive jets are calculated in both p+p and Pb+Pb collisions  at $\rm \sqrt{s}=5.02$~TeV at three energy intervals: $\rm 5<E_\text{Radiator}<10$~GeV, $\rm 10<E_\text{Radiator}<20$~GeV and  $\rm 20<E_\text{Radiator}<35$~GeV respectively.}
    \label{tab:avertojets}
\end{table}
\begin{table}
    \centering
    \begin{tabular}{|c|c|c|c|c|c|}
        \hline
        \multirow{2}{*}{ $E_\text{Radiator}$} &   \multicolumn{2}{c|}{Inclusive jets}    &   \multicolumn{2}{c|}{$D^0$ jets}        & \multirow{2}{*}{ } \\
        \cline{2-5}
        & $\rm \langle \theta \rangle_{spl} $ & $\rm  N_{spl}$ &$\rm \langle \theta \rangle_{spl}$  & $\rm N_{spl}$ & \\
        \hline
         \multirow{2}{*}{$5-10$~GeV} & 0.227 & 1.358& 0.277 & 1.233  & pp \\
        \cline{2-6}
         & 0.256 & 1.405 & 0.280  & 1.280  &AA\\
         \hline
          \multirow{2}{*}{$10-20$~GeV} & 0.220 &  1.810 & 0.244  & 1.510 & pp \\
        \cline{2-6}
         & 0.254 & 1.757 & 0.263  & 1.600  &AA\\
         \hline
           \multirow{2}{*}{$20-35$~GeV}& 0.232 &  2.040   & 0.232 &  1.822  & pp \\
        \cline{2-6}
         & 0.249  & 1.977    & 0.251 &  1.860   &AA\\
         \hline
    \end{tabular}
    \caption{The averaged splitting-angles per splitting for $\rm D^0$-meson tagged and for inclusive jets are calculated in both p+p and Pb+Pb collisions at $\rm \sqrt{s}=5.02$~TeV at three energy intervals: $\rm 5<E_\text{Radiator}<10$~GeV, $\rm 10<E_\text{Radiator}<20$~GeV and  $\rm 20<E_\text{Radiator}<35$~GeV. The numbers of splittings in jets are also provided accordingly.}
    \label{tab:avertospli}
\end{table}

We also compute in the Fig.~4 the splitting-angle distributions for inclusive jets normalized to the number of jets in Pb+Pb collisions at $\rm \sqrt{s}=5.02$~TeV and also the A+A/p+p ratios for all three energy intervals of the radiators. We find, both in the light flavor and heavy flavor case, the splitting-angle distributions are always shifting to larger $\theta$ (small $\rm \ln(1/\theta)$) due to jet quenching. We can easily summarize that the medium-induced radiation will broaden the splittings for heavy and light-initiated parton compared to the case in p+p.

When taking a closer look at the value of heavy/light ratios demonstrated in Fig.~1 and Fig.~2, unlike what we observed in A+A/p+p ratios discussed above, we find the number of jets normalized emission-angle distribution ratios in p+p are smaller than  $1$ at all the investigated $\theta$ region at $\rm 20<E_\text{Radiator}<35$~GeV, in the case of A+A, the ratios are below $1$ in $\rm 10<E_\text{Radiator}<20$~GeV and $\rm 20<E_\text{Radiator}<35$~GeV. It is not easy to understand such suppression at all the investigated. To address such concern, we systematically calculate the averaged values of splitting angles per jets in $\rm D^0$-meson (inclusive) jets in p+p and Pb+Pb at $\rm \sqrt{s}=5.02$~TeV demonstrated in Table.~1 for each energy interval. Indeed, we find the averaged values for $\rm D^0$ jets are smaller than inclusive ones both in p+p and A+A in $\rm 10<E_\text{Radiator}<20$~GeV and $\rm 20<E_\text{Radiator}<35$~GeV which is consistent with what we observed in the heavy/light ratios as functions of $\theta$. It is pointing to an important caveat.

We believe the number of splittings along the tracked prong in the investigated jets needs to be considered and further investigated. In the Table.~\ref{tab:avertospli}, averaged splitting angles per splitting are calculated in $\rm D^0$-meson (inclusive) jets in p+p and Pb+Pb at $\rm \sqrt{s}=5.02$~TeV, besides that, the numbers of reconstructed splittings in jets are also provided accordingly. We find the averaged splitting angles per splitting of the $\rm D^0$-meson tagged jets are larger than that of the inclusive jets both in p+p and A+A, and the averaged splitting angles per splitting in A+A are also larger than their counterparts in p+p. We can also find the numbers of splittings in $\rm D^0$-meson tagged jets in each energy interval is always smaller than that in inclusive jets.  All the calculations are cross-checked with LBT calculations~\cite{He:2015pra,Cao:2016gvr}, The values of the calculation results in A+A collisions using LBT are $1 -2\%$ larger than that by using SHELL. Then, the scenario of the dead-cone effect is that the probability of heavy quark emitting gluon at a smaller angle is largely suppressed due to the dead-cone effect. It will lead to the emissions being distributed at a larger angle. However, the possibility of emitting such gluon is suppressed due to this mass effect governed by the dead-cone term in Eq.~\ref{eq:dNdxdk2}.

\section{Conclusions}

Using the SHELL model, we have calculated the emission angle distribution of charm-quark initiated splittings and that of light parton initiated splittings, both in pp and Pb+Pb at $\rm \sqrt{s}=5.02$~TeV respectively. First, we have validated our vacuum reference calculation by comparing it to the ALICE p+p dead cone results. We then report the results of the full simulation, which includes medium-induced radiation as well as collisional energy loss. We find that the splitting angle distributions get broader in Pb+Pb relative to p+p due to medium-induced radiation both for $\rm D^0$-meson tagged jets and inclusive jets. The mass effect will also induce the splitting angle distributions getting broader both in p+p and Pb+Pb. Still, we observe dead-cone signal in PbPb: a strong suppression of small-angle splittings for $\rm D^0$-meson tagged jets relative to inclusive ones. We also find the collisional energy loss does not induce suppression of small splitting angles for $\rm D^0$-meson tagged jets. The scenario of the dead-cone effect is that the survived splitting angle of heavy flavor-initiated splitting is distributed at a larger angle, however, the possibility of such emission will be reduced.


{\bf Acknowledgments:} 
The authors would like to thank X Peng for helpful discussions. This research is supported by the Guangdong Major Project of Basic and Applied Basic Research No.2020B0301030008, the Natural Science Foundation of China with Project Nos. 11935007 and 11805167.


\vspace*{-.6cm}

\end{document}